# Tuning of envelope in high overtone bulk acoustic wave resonator

Kongbrailatpam Sandeep, J Pundareekam Goud, and K. C. James Raju [a)]

*CASEST, School of Physics, University of Hyderabad, Hyderabad, 500019, India*

This report presents a robust composite resonator known as the high overtone bulk acoustic wave resonator (HBAR) for demonstrating experimentally the ability of the resonator to tune the envelope of the spectrum by applying dc bias. The reported HBAR exhibits exceptionally high-quality factor in the frequency range of 700 MHz to 3 GHz and high effective coupling coefficient over the broad range of microwave frequencies. The HBAR is based on thin $Ba_{0.5}Sr_{0.5}TiO_3$ film sandwiched between two electrodes forming a transducer, supported by a thick sapphire substrate which is a low acoustic loss material. The resonator works on the principle of induced piezoelectricity due to the applied dc bias. The minima of the envelope changes from 1.16 GHz with $S_{11}$ of -3.51 dB at 100 kV/cm bias to 1.90 GHz with $S_{11}$ of -42.23 dB at 700 kV/cm bias, giving a relative tunability of the envelope to be around 64%. The spacing of the parallel resonance frequency (SPRF) for different bias voltages are also presented in this report. The quality factor of the resonator is over 22,000 at around 2GHz.

High overtone bulk acoustic wave resonators (HBAR) have been used for various applications but most prominently in material characterization and oscillators[1]. Certain aspects of HBAR's applicability in gas sensing and gravimetric sensing have also been explored [2,3]. The applicability of HBAR when compared to film bulk acoustic wave resonator (FBAR) and solidly mounted resonator (SMR) are quite different even though they fundamentally employ either piezoelectric effect or induced piezoelectric effects. The use of paraelectric phase ferroelectric thin film like $Ba_{0.5}Sr_{0.5}TiO_3$ has opened up a whole new paradigm in the field of acoustic wave devices employed in filtering applications. Utilization of induced piezoelectricity in such devices have made it possible to make switchable and tunable filters for various application which reduces switches and filters in the RF front end [4-6]. The use of symmetric and asymmetric displacement modes by utilizing multilayers of thin films have also been reported by other

[a)] Author to whom correspondence should be addressed: kcjrsp@uohyd.ernet.in

groups [7]. In our previous work we have reported a switchable HBAR based on BST thin film and by using resonant spectrum method, the BST thin film was characterized [1,8].

In this work, we fabricate and characterize a HBAR based on BST thin film with very high effective coupling coefficient and high Q factor over a broad microwave frequency range. We also demonstrate experimentally the tuning of the envelope of HBAR by application of bias. The HBAR structure can be primarily considered as having two components, the first is the transducer which is composed of the thin BST film sandwiched by a top and a bottom electrode and the second is the thick plane parallel, and low acoustic loss substrate like sapphire. The transducer of the resonator is from where the acoustic wave is generated, and the envelope of the spectrum is dependent on the various aspects of the transducer and its active layer BST. The substrate stores the energy transferred from the transducer and it heavily affects the spacing of parallel resonance frequency (SPRF).

A c-axis oriented sapphire substrate of 4-inch diameter and thickness 500 μm is taken and cut into dies of 1cm x 1 cm. The substrate is then coated with Platinum (Pt) using RF sputtering technique with a thickness of around 120 nm. Then, BST of thickness 1.13 μm is coated using Pulse Laser Deposition technique [1]. Gold (Au) of thickness around 150 nm is deposited above the BST layer, and the top electrode which has the circular patch capacitor (CPC) structure is obtained by using photolithographic techniques and lift-off process. The microwave measurement set-up consists of Agilent E8361C network analyzer, G-S-G 250 μm probes (J microTechnology), on wafer probe station, calibration substrates, bias tee and a voltage source.

Measurements are performed for the frequency range of 200 MHz to 3.5 GHz, and different bias are given to the resonator starting from 100 kV/cm to 700 kV/cm. Figure 1 is the plot of the scattering parameter, $S_{11}$(dB) versus the frequency. Figure shows the resonant modes and the envelope in frequency spectrum of the resonator. There is a change in the intensity of the peaks when the bias applied to the resonator is increased gradually. The resonances occur in the spectrum due to induced piezoelectric effect by application of dc bias in the paraelectric phase of thin film BST. The HBAR exhibits multiple resonant modes spanning a broad range of microwave frequency because of the thick sapphire substrates and due to creation of standing waves. The low acoustic loss substrate couples with the transducer and traps energy inside its volume.

Apart from the above-mentioned effect in the resonator, another interesting observation is the switching or hopping of modes from the lower to higher frequency when the bias voltage increases progressively. The switching of two modes has already been discussed by Vorobeiv et al by using symmetric and asymmetric displacement modes by utilizing multilayers of thin films, where, by changing the polarity, the modes are switched [7]. In our work, the applied bias is limited to applying just one polarity, but the observed hopping is quite different to what was reported in [7]. Table I give a detailed interpretation of the plots in Fig. 1 and specify the minima in the envelope at different frequencies. It can be inferred from the table that, like the frequency tuning in FBAR and SMR, HBAR can also be tuned via its envelope which depend significantly on the active ferroelectric thin BST layer. Such technique of switching and tuning the resonance mode over a broad frequency range is especially interesting as $k_{eff}^2$ and Q factor of the resonator is very high when compared to other acoustic wave devices. The relative tunability percentage obtained when calculated for a change in bias of 100 KV/cm and 700 KV/cm is around 64 %.

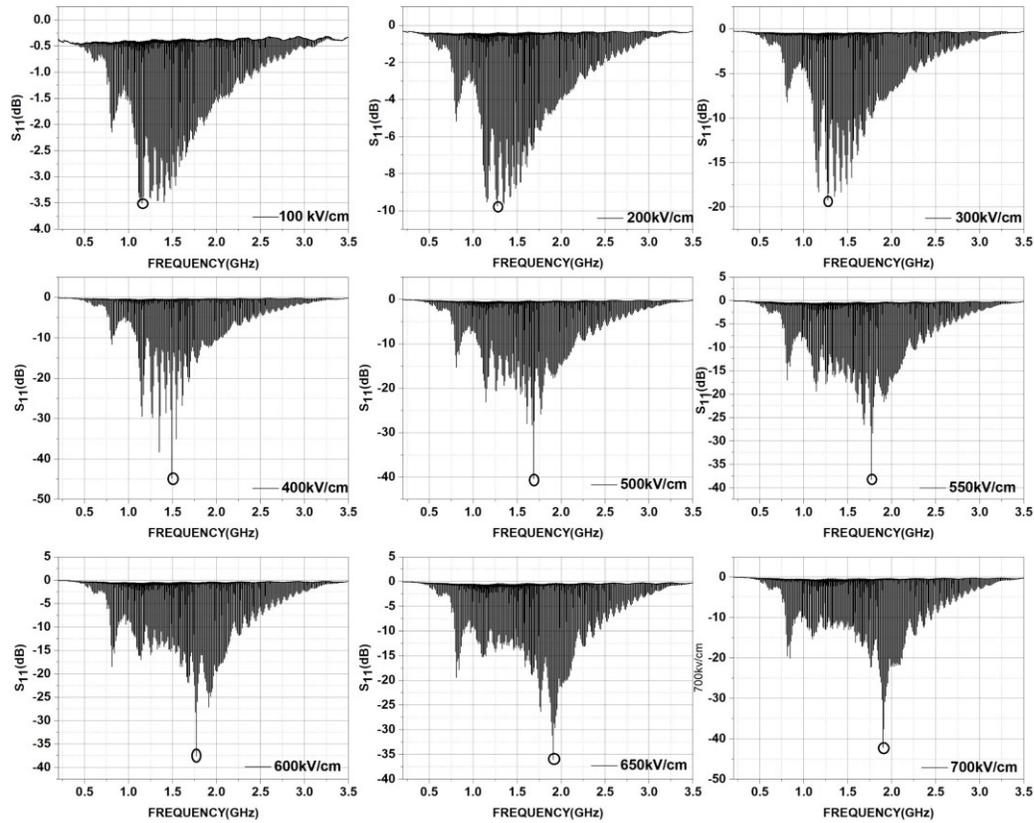

Figure 1: Frequency spectra of HBAR at various bias(circle in inset specifies minima).

Table 1: Minima of the envelope.

| Bias field, kV/cm | 100 | 200 | 300 | 400 | 500 | 550 | 600 | 650 | 700 |
|---|---|---|---|---|---|---|---|---|---|
| Minima Freq,GHz | 1.16 | 1.28 | 1.28 | 1.49 | 1.68 | 1.77 | 1.77 | 1.91 | 1.90 |

Figure 2 show the $k_{eff}^2$ and Q factor of the resonator at various bias voltages. The resonator exhibits exceptionally high Q factor of over 20,000 around 2GHz. The figure of merit (FOM) is 4 x 10 [13] for this resonator and can be used for oscillator applications. The coupling coefficient value is also impressive for application in filter designs. Figure 3 show the spacing of parallel resonance frequency (SPRF). Since the SPRF of the modes in HBAR depend significantly on the substrate properties there is no significant change or variation by application of bias. Thus, we can confirm that the envelope hopping effect is caused by the bias on the metal insulator metal (MIM) structure with the active BST layer. The same follows for $k^2_{eff}$ and Q factor for the HBAR except the SPRF distribution.

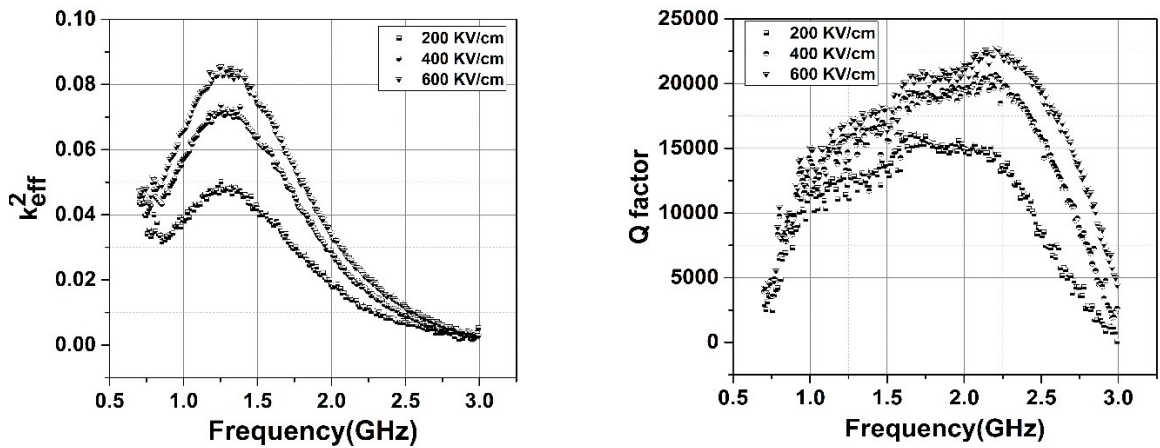

Figure 2: $k^2_{eff}$ and Q factor distribution of HBAR modes at various bias voltages.

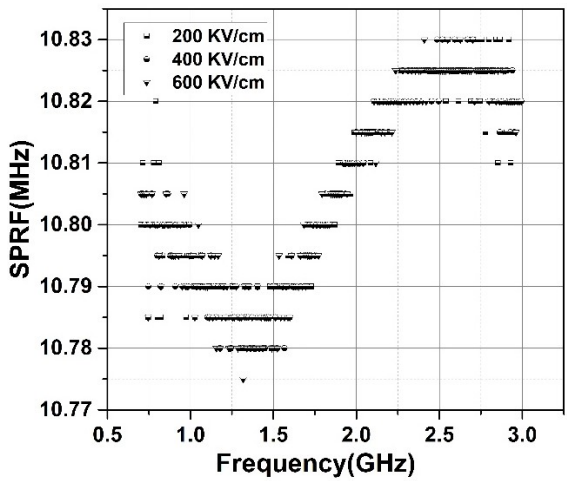

Figure 3: SPRF distribution of HBAR at various bias voltages.

In this letter, the switching and hopping nature of the resonant modes of the HBAR is experimentally demonstrated. And the relative tuning of the envelope of the frequency spectra of the HBAR is around 64 %. The HBAR fabricated also exhibits very high Q factor and coupling coefficient in the GHz range of frequency. Such resonators occupy immense potential in the field of filters and oscillator design as the figure of merit (FOM) is 4 x 10[13]. The $k_{eff}^2$ and Q factor for the HBAR is highly bias dependent while the SPRF is not.